\documentclass[11pt,a4paper,aps,showpacs,superscriptaddress,nofootinbib]{revtex4-1}
\usepackage[latin9]{inputenc}
\usepackage{color}
\usepackage{amsmath}
\usepackage{amssymb}

\makeatletter

\pdfpageheight\paperheight
\pdfpagewidth\paperwidth

\providecommand{\tabularnewline}{\\}

\usepackage{upgreek}

\def\cep{\ep} 
\def\al{\alpha}
\def\be{\beta}
\def\ga{\gamma}

\def\ep{\epsilon}

\def\et{\eta}

\def\ka{\kappa}
\def\la{\lambda}

\def\si{\sigma}

\def\ch{\chi}
\def\ps{\psi}
\def\om{\omega}
\def\Ga{\Gamma}

\def\cl{{\cal L}}

\def\mn{{\mu\nu}}

\def\half{{\textstyle{1\over 2}}}

\def\frac#1#2{{\textstyle{{#1}\over {#2}}}}

\def\lsim{\mathrel{\rlap{\lower4pt\hbox{\hskip1pt$\sim$}}
    \raise1pt\hbox{$<$}}}
\def\gsim{\mathrel{\rlap{\lower4pt\hbox{\hskip1pt$\sim$}}
    \raise1pt\hbox{$>$}}}
\def\sqr#1#2{{\vcenter{\vbox{\hrule height.#2pt
         \hbox{\vrule width.#2pt height#1pt \kern#1pt
         \vrule width.#2pt}
         \hrule height.#2pt}}}}

\def\prt{\partial}
\def\lrpartial{\raise 1pt\hbox{$\stackrel\leftrightarrow\partial$}}

\def\lrDmu{\stackrel{\leftrightarrow}{D_\mu}}

\def\Re{\hbox{Re}\,}
\def\Im{\hbox{Im}\,}

\def\pt#1{\phantom{#1}}

\def\f{$f_\mu$}

\def\lrDmu{{\hskip -3 pt}\stackrel{\leftrightarrow}{D_\mu}{\hskip -2pt}}
\newcommand{\beq}{\begin{equation}}
\newcommand{\eeq}{\end{equation}}
\newcommand{\bea}{\begin{eqnarray}}
\newcommand{\eea}{\end{eqnarray}}
\def\nsc#1#2#3{\om_{#1}^{{\pt{#1}}#2#3}}
\def\lsc#1#2#3{\om_{#1#2#3}}

\def\lulsc#1#2#3{\om_{#1\pt{#2}#3}^{{\pt{#1}}#2}}

\def\tor#1#2#3{T^{#1}_{{\pt{#1}}#2#3}}

\def\vb#1#2{e_{#1}^{{\pt{#1}}#2}}
\def\ivb#1#2{e^{#1}_{{\pt{#1}}#2}}
\def\uvb#1#2{e^{#1#2}}
\def\lvb#1#2{e_{#1#2}}

\makeatother

\begin{document}

\title{The role of singular spinor fields in a torsional gravity, Lorentz-violating,
framework}

\author{A. F. Ferrari}
\email{alysson.ferrari@ufabc.edu.br}

\affiliation{Universidade Federal do ABC - UFABC, 09210-580, Santo André, SP,
Brazil}

\author{J. A. S. Neto}
\email{jose.antonio@ufabc.edu.br}

\affiliation{Universidade Federal do ABC - UFABC, 09210-580, Santo André, SP,
Brazil}

\author{R. da Rocha}
\email{roldao.rocha@ufabc.edu.br}

\affiliation{Universidade Federal do ABC - UFABC, 09210-580, Santo André, SP,
Brazil}

\begin{abstract}
In this work, we consider a generalization of quantum electrodynamics
including Lorentz violation and torsional-gravity, in the context
of general spinor fields as classified in the Lounesto scheme. Singular
spinor fields will be shown to be less sensitive to the Lorentz violation,
as far as couplings between the spinor bilinear covariants and torsion
are regarded. In addition, we prove that flagpole spinor fields do
not admit minimal coupling to the torsion. In general, mass dimension
four couplings are deeply affected when singular \textendash{} flagpole 
\textendash{} spinors are considered, instead of the usual Dirac spinors.
We also construct a mapping between spinors in the covariant framework
and spinors in Lorentz symmetry breaking scenarios, showing how one
may transliterate spinors of different classes between the two cases.
Specific examples concerning the mapping of Dirac spinor fields in
Lorentz violating scenarios into flagpole and flag-dipole spinors
with full Lorentz invariance (including the cases of Weyl and Majorana
spinors) are worked out. 
\end{abstract}
\maketitle

\section{Introduction}

Quantum field theory and particle physics demand that matter is described
by fermionic fields. At a first level of classification, previously
to second quantization, this means that spinor fields are essential
to describe physics. Although Dirac spinors are the most usual objects
that carry spinor representations of the Lorentz group, they are actually
just the tip of the iceberg that encompasses a comprehensive set of
possibilities, described by the Lounesto classification\,\cite{lou2,oxford},
based upon the bilinear covariants. Dirac spinors are included as
particular cases of regular spinors, while Majorana and Weyl spinors
are well-known examples in the classes of flagpole and dipole spinors,
respectively. Besides these, the Lounesto classification also describes
a huge class of new possibilities, including mass dimension one spinors\,\cite{daSilva:2012wp,cyleeI,Lee:2014opa,EPJJC},
exotic formulations with dynamical mass generation mechanisms \cite{Bernardini:2012sc},
and solutions of the Dirac equations in specific circumstances, which
are not Dirac spinors. There are still sub-classes
in the Lounesto classification which remain unexplored, whose dynamics
are still unknown\,\cite{daSilva:2012wp,cyleeI,EPJJC}. Reciprocal,
equivalent, classifications have further paved recent developments
\cite{oxford,Cavalcanti:2014wia,fabbriij}. 
Many efforts have been devoted to unravel those less known spinors
classes. Flagpole spinor include for example Elko and Majorana spinors,
and have been used in different contexts, from particle physics and
LHC phenomenology to cosmology. Elko spinors have the peculiar feature
of being mass dimension one spinor fields (they are not the only ones,
as recently pointed out in\,\cite{EPJJC}). As some recent studies
regarding flagpole spinors, we can quote those involving: possible
signatures via monojets at 14 TeV in LHC and as a byproduct of a Higgs
bubble\,\cite{Dias:2010aa}, tunnelling methods and Hawking radiation\,\cite{daRocha:2014dla},
as well as some cosmological aspects\,\cite{Fabbri:2011mi,Fabbri:2012qr,Fabbri:2010pk,Fabbri:2013gza,saulo}.

A thorough analysis on the role of flagpoles and flag-dipole spinors
and their interpretation in the Penrose formalism was derived in Refs.
\cite{lou2,daSilva:2012wp}. New spinors solutions in supergravity
have been also recently obtained \cite{bonora,Bonora:2015ppa}, respectively, in the 
context of the AdS$_4 \times S^7$ and AdS$_5 \times S^5$ compactifications. As
a major physical motivation, flagpole spinor fields have been explored
as candidates for dark matter\,\cite{cyleeI,Lee:2014opa}.

Flag-dipoles are spinor fields that were not listed in physics applications
until recently, when this class of spinors was shown to provide a
solution of the Dirac equation in a $f(R)$, torsional, gravity setup\,\cite{riemanncartan}.
In fact, in ESK gravity, torsion is coupled to the spin density of
the matter field. Therefore, all terms involving the covariant derivatives
and the curvature can be split into their torsionless counterparts
plus torsional contributions, which can be replaced by the torsion-spin
field equations in terms of the spin density of the spinor matter
field \cite{riemanncartan,Fabbri:2011mi,Fabbri:2012qr,Fabbri:2010pk}.
ESK theory is equivalent to a theory complemented by spin-spin self-interacting
potentials, with non-linearities in the matter field equations. The
specific gravitational background, for example $f(R)$ or conformal,
and the type of spinor (Dirac or other regular spinors; flag-dipoles,
flagpoles, or dipole spinors \textendash{} the singular ones \cite{daSilva:2012wp})
shall determine the exact structure of these non-linearities in the
matter field equations. For example, in the least order derivative
ESK gravity with Dirac fermion fields, the non-linearities are realized
by axial current squared contact interactions, provided by Nambu-Jona-Lasinio
(NJL) potentials. When the spinor field is a flag-dipole fermion field,
the interaction is shown to be changed\,\cite{riemanncartan}.

On the other hand, Lorentz symmetry is the fundamental symmetry underlying
the standard model (SM) of particle physics, being experimentally
already verified at exquisite level of precision. Nevertheless, the
paradigm of Lorentz symmetry might be modified in very high energy
regimes. Indeed, whenever quantum gravitational effects can not be
dismissed, Lorentz symmetry must be malleable to adequate these extremal
regimes into physics. As an example, Lorentz symmetry might be spontaneously
broken by some tensorial fields acquiring non-vanishing vacuum expectation
values in the low energy effective theory of string theory\,\cite{Kostelecky:2007kx}.
A rather general approach in which Lorentz symmetry violation (LV)
is incorporated within the Standard Model was developed by Colladay
and Kostelecky\,\cite{collakoste,colladay4}. This setup has been
used as a framework for studying consequences of LV in a plethora
of physical phenomena, imposing very stringent bounds on the LV parameters\,\cite{Kostelecky:2007kx,Bernardini:2007uj,Charneski:2012py,Maluf:2015hda}.
Some LV phenomena that have recently received attention in the literature
are effects in acoustic black holes\,\cite{anacleto1}, BPS vortices
in a background with Lorentz and CPT breaking\,\cite{manoel2,Bernardini:2007ez},
thick branes in LV\,\cite{menezes}, among many others.

Our main goal is to study the role of LV in theories containing singular
spinor fields in Riemann-Cartan spacetimes. Since the Palatini and
the Einstein-Hilbert Lagrangian densities are physically equivalent,
also in the context of the Lounesto classification of spinors, 
curvature and torsion are equivalent descriptions of the gravitational
field. The matter energy-momentum tensor is the source of curvature,
in the case of general relativity, and of torsion, in the case of
teleparallel gravity. Moreover, torsion-spin coupling can be regarded,
besides the curvature-energy coupling. In addition, background torsion
violates effective local Lorentz invariance\,\cite{Kostelecky:2007kx}.
Riemann-Cartan geometry is the setup for the Einstein-Cartan theory,
however, more general gravitation theories in Riemann-Cartan spacetimes
may incorporate propagating \emph{vierbein} and spin-connection fields,
describing dynamical torsion and curvature\,\cite{waldyr}.

To consider the torsion coupling to matter fields, usually Dirac fermion
fields are employed in the literature. In this setup, torsion effects
manifest as self-interactions, capable of providing a dynamical explication
of the exclusion principle\,\cite{Fabbri:2012qr}. Gauge couplings
on this setup was considered in Ref. \cite{Casana:2002fu} and its
ramifications. Torsion effects
influence the dynamics near or at the Planck scale, wherein breaking
of Lorentz invariance could be relevant. Moreover, torsion induces
a preferred orientation for a freely falling observer, realized as
a manifestation of local Lorentz violation. Hence, constraints on
Lorentz symmetry have led to constraints on torsion\,\cite{Kostelecky:2007kx}.
We want to analyze the torsion coupling to matter fields beyond Dirac
fermion fields, encompassing in the analysis the whole set of singular
spinor fields, in particular flagpole and flag-dipole spinor fields.
One of our motivations is that recently the Riemann-Cartan geometry
shed new light on the prominent roles of singular spinors: it was
shown that the Einstein-Sciama-Kibble theory coupled to spinors admits
solutions that are not Dirac spinor fields, but flag-dipoles ones\,\cite{riemanncartan}.

Regardless of its origin in some high energy theory, Lorentz violation
is usually accomplished in an effective field theory framework. The
so-called Standard Model Extension (SME) \cite{colladay4} contains
General Relativity (GR) and the Standard Model (SM). Dominant terms
in the SME action include the pure gravity and minimally coupled SM
actions, together with all leading-order terms introducing Lorentz
symmetry violations that can be constructed from gravitational and
SM fields, coupled to constant background tensors. The Riemann-Cartan
geometry allows for non-zero vacuum quantities that violate local
Lorentz invariance, although preserving general coordinate invariance,
encompassing minimal gravitational couplings of spinors. We will show
that, in this context, flagpole spinor fields are examples of singular
spinor fields that can not couple minimally to torsion.

It is also well known that not all LV parameters in the SME are physically
meaningful\,\cite{collakoste}, in the sense that some of them may
be absorbed by a field redefinition. Moreover, a properly redefined
conserved current satisfies the usual Poincaré algebra, at least as
far as these coefficients are concerned. This point was extensively
worked out in\,\cite{colladay2}, where a systematic procedure for
eliminating spurious LV coefficients and defining conserved currents
was developed for the case of QED. The implications of this procedure
for theories involving more general types of spinor fields is still
lacking, and this is another question we address in this work.

This paper is organized as follows: in Sect.\,\ref{sec:Coupling-singular-spinors}
we outline the main setup, briefly introducing the Lounesto classification
and essentials on LV and the Riemann-Cartan framework. We shall argue
that couplings between the spinor bilinear covariants and torsion
are an appropriate signature to probe the sensitiveness of singular
spinor fields to LV parameters. We prove that flagpole spinor fields
do not admit minimal coupling with torsion, and couplings with mass
dimension four coefficients are influenced by the choice of flagpoles
spinors, instead of the usual Dirac spinors. In Sect.\,\ref{sec:A-bridge-between},
the flat spacetime is regarded, and discuss a general mapping between
different classes of spinors in LV and Lorentz covariant theories.
In particular, we relate regular, Dirac spinors, in a Lorentz violating
theory, into singular, flagpole and flag-dipole spinors, in a full
Lorentz invariant framework. In Sect.\,\ref{sec:Concluding-remarks-and}
the conclusions are presented, further analyzing outlooks and perspectives.

\section{\label{sec:Coupling-singular-spinors}Coupling singular spinors to
torsion in Lorentz violating framework}

Let us consider the set of spinor fields in Minkowski spacetime $M\simeq\mathbb{R}^{1,3}$.
Given sections of the frame bundle ${P}_{\mathrm{SO}_{1,3}^{e}}(M)$
$\{{\rm e}_{\mu}\}$, with dual basis $\{{\rm e}^{\mu}\}$, classical
spinor fields carrying a ${(1/2,0)}\oplus{(0,1/2)}$ representation
of the Lorentz group component connected to the identity $\mathrm{SL}(2,\mathbb{C)}\simeq\mathrm{Spin}_{1,3}^{e}$
are sections of the bundle ${P}_{\mathrm{Spin}_{1,3}^{e}}(M)\times\mathbb{C}^{4}.$
Denoting by $\{\gamma^{\mu}\}$ the set of gamma matrices, the bilinear
covariants are given by 
\begin{align}
\sigma & =\bar{\psi}\psi,\nonumber \\
\mathbf{J} & =J_{\mu}{\rm e}^{\mu}=\bar{\psi}\gamma_{\mu}\psi{\rm e}^{\mu},\nonumber \\
\mathbf{S} & =S_{\mu\nu}\;{\rm e}^{\mu}\wedge{\rm e}^{\nu}=\frac{1}{2}\bar{\psi}i\gamma_{\mu\nu}\psi\;{\rm e}^{\mu}\wedge{\rm e}^{\nu},\nonumber \\
\mathbf{K} & =K_{\mu}\;{\rm e}^{\mu}=\bar{\psi}\gamma_{5}\gamma_{\mu}\psi\;{\rm e}^{\mu},\nonumber \\
\omega & =\bar{\psi}\gamma_{5}\psi\thinspace,\label{fierz}
\end{align}
where $i\gamma^{5}=\gamma^{0}\gamma^{1}\gamma^{2}\gamma^{3}$. Exclusively
in the Dirac theory of the electron, $\mathbf{J}$ is interpreted
as being the current density, $\mathbf{K}$ provides the direction
of the electron spin, and $\mathbf{S}$ is related to the distribution
of intrinsic angular momentum. This physical interpretation is absent
in the most general cases. Whenever $\omega=0=\sigma$, the spinor
field is said to be singular, otherwise it is a regular spinor. The
bilinear covariants for regular spinors satisfy the Fierz identities,
\begin{eqnarray}\label{fffie}
\mathbf{K}\cdot\mathbf{J}=0\,,\qquad\mathbf{S}=(\omega+\sigma i\gamma_{5})^{-1}\mathbf{K}\wedge\mathbf{J}\,,\qquad\mathbf{J}^{2}=\omega^{2}+\sigma^{2}=-\mathbf{K}^{2}\,.
\end{eqnarray}

Lounesto spinor field classification is based on six disjoint spinor
field classes~\cite{lou2}, 
\begin{eqnarray}
1)\;\;\sigma\neq0,\;\;\;\omega\neq0,\;\;\;\mathbf{J}\neq0\qquad{}\qquad{}\qquad{}\qquad{}\qquad{}4)\;\;\sigma=0=\omega,\;\;\;\mathbf{S}\neq0,\;\;\;\mathbf{K}\neq0,\;\;\;\mathbf{J}\neq0\label{Elko11}\\
2)\;\;\sigma\neq0,\;\;\;\omega=0,\;\;\;\mathbf{J}\neq0\qquad{}\qquad{}\qquad{}\qquad{}\qquad{}5)\;\;\sigma=0=\omega,\;\;\;\mathbf{S}\neq0,\;\;\;\mathbf{K}=0,\;\;\;\mathbf{J}\neq0\label{tipo41}\\
\!\!\!3)\;\;\sigma=0,\;\;\;\omega\neq0,\;\;\;\mathbf{J}\neq0\qquad{}\qquad{}\qquad{}\qquad{}\qquad{}\!6)\;\;\sigma=0=\omega,\;\;\;\mathbf{S}=0,\;\;\;\mathbf{K}\neq0,\;\;\;\mathbf{J}\neq0\label{dirac21}
\end{eqnarray}
The first three types of spinor fields, characterized by $\mathbf{J},\mathbf{K},\mathbf{S}\neq0$,
are regular. The last three are known, respectively, as flag-dipole,
flagpole and dipole spinor fields. It is worth to mention that the
paradigm of Lounesto classification is concerned with ${\bf J}\neq0$.
The mechanism to generate three additional classes has been proposed
in\,\cite{EPJJC}, including the case ${\bf J}=0$, implying that
the subsequent spinors have canonical mass dimension one, being conjectured
to be ghost fields.

At this point, we review how LV can be introduced in a general way
into a theory such as QED, and how some of the LV thus introduced
is actually spurious, being eliminated by a proper field redefinition.
We shall be interested only in the free fermion piece of the QED Lagrangian,
which in the LV case is written as 
\begin{align}
\mathcal{L}^{\text{LV-QED}} & ={\frac{i}{2}\bar{\psi}\Gamma^{\nu}\overset{\leftrightarrow}{\nabla_{\nu}}\psi}-\bar{\psi}M\psi\thinspace,\label{eleclagran}
\end{align}
where $\nabla_{\mu}=\partial_{\mu}+iqA_{\mu}$ is the covariant derivative,
and 
\begin{eqnarray}
\Gamma_{\nu} & := & \gamma_{\nu}+c_{\mu\nu}\gamma^{\mu}+d_{\mu\nu}\gamma_{5}\gamma^{\mu}+{\mathfrak{e}_{\nu}}+if_{\nu}\gamma_{5}+\frac{1}{2}g_{\rho\mu\nu}\sigma^{\rho\mu}\,,\label{gamaLV}\\
M & := & m+im_{5}\gamma_{5}+a_{\mu}\gamma^{\mu}+b_{\mu}\gamma_{5}\gamma^{\mu}+\frac{1}{2}H_{\mu\nu}\sigma^{\mu\nu}\thinspace,\label{mmm}
\end{eqnarray}
$m$ being the electron mass, $\sigma^{\mu\nu}=\frac{i}{2}[\gamma^{\mu},\gamma^{\nu}]$,
and $a,b,c,d, {\mathfrak{e}},f,g,m_{5},H$ are real
constant tensors which parametrize the LV. By assumption, the action\,\eqref{eleclagran}
is Hermitian, hence constraining the coefficients for Lorentz violation
to be real. Some of these parameters have a strong experimental/phenomenological
bound, as discussed in\,\cite{Kostelecky:2007kx}. However, some
parameters in Eqs.\,\eqref{gamaLV} and\,\eqref{mmm} can be eliminated
by a proper field redefinition.

Indeed, a spinor $\chi$ that satisfies a LV extended QED Lagrangian
can be obtained from a spinor $\psi$, which is a solution of the
standard Lorentz invariant (LI) QED Lagrangian, through the transformation
{
\begin{eqnarray}
\psi & =(\mathbf{1}+f(x^{\mu},\partial_{\nu}))\chi=\chi+(v\cdot\Gamma+i{\rm e}+i\tilde{C}_{\mu}x^{\mu}+C_{\mu\nu}x^{\mu}\partial^{\nu}+B_{\mu}\partial^{\mu}+\gamma_{5}\tilde{B}_{\mu}\partial^{\mu})\chi\thinspace,\label{spintrans}
\end{eqnarray}
}where $f(x,\partial)$ represents a general $4\times4$ matrix function
of the coordinates and derivatives\,\cite{colladay2}. Here, $v\cdot\Gamma=v_{I}\Gamma_{I}$,
$\Gamma_{I}$ being a basis for ${\cal {M}}(4,\mathbb{C})$ for a
composed index $I\in\{{\scriptstyle \emptyset,\mu,\mu\nu,\mu\nu\rho,{5}}\}$,
where $\Gamma_{\emptyset}=\boldsymbol{1}$. In addition, ${\theta},\thinspace{\tilde{C}_{\mu}},\thinspace B_{\mu},\thinspace\tilde{B}_{\mu},\thinspace C_{\mu\nu}$
are scalar coefficients. The parameters $\Re({\theta})$,
$B_{\mu}$ and $C_{[\mu\nu]}$ in particular correspond to the $U(1)$
and Poincaré symmetries of the standard Lagrangian. Only lowest order
terms in the field redefinition are retained, since the LV parameters
are assumed to be small. These redefinitions can be regarded as position-dependent
mixings of components in spinor space.

For simplicity, we quote the result of the redefinition parametrized
by the parameter $v$, which have been first described in\,\cite{collakoste}.
We start with an explicitly LI Lagrangian 
\begin{equation}
\mathcal{L}=\frac{i}{2}\bar{\psi}\gamma^{\mu}\stackrel{\leftrightarrow}{\partial_{\mu}}\psi-m\bar{\psi}\psi\thinspace,\label{eq:LLI}
\end{equation}
which is rewritten via the redefinition\,\eqref{spintrans} as 
\begin{align}
\mathcal{L} & =\frac{i}{2}\bar{\chi}\gamma^{\mu}\stackrel{\leftrightarrow}{\partial_{\mu}}\chi-m\bar{\chi}\chi+\frac{i}{2}\bar{\chi}[\{\gamma^{\mu},\Gamma\cdot{\rm \Re}v\}+i[\gamma^{\mu},\Gamma\cdot{\rm \Im}v]]\stackrel{\leftrightarrow}{\partial_{\mu}}\chi-2m{\rm \Re}v\cdot\bar{\chi}\Gamma\chi\thinspace.\label{eq:LLV}
\end{align}
The essential point is that both Lagrangians describe the same physics,
so even if Eq.\,\eqref{eq:LLV} includes terms that might seem to
violate Lorentz invariance, this theory is actually Lorentz invariant.
Indeed, one can define properly modified Poincaré generators in terms
of $\chi$ that satisfy the usual Poincaré algebra. This example shows
that the general field redefinition described by Eq.\,\eqref{spintrans}
can eliminate some of the coefficients in a general LV effective theory
as the SME, since these coefficients are actually unobservable at
leading order in SME-related phenomenology.

It is worth to emphasize that Lagrangians having derivative terms for spinors are here adopted. Hence, the field equations have terms given by the Dirac operator and terms that have a similar differential order that influence the dynamics and the light-cone structure as well \cite{VeloZwan1,VeloZwan2}. 
Interactions that have the same derivative order of the leading kinetic term can yield pathologies like the mismatch between degrees of freedom and field equations and superluminal propagation, among others. In general, matter fields are classified according to their spin \cite{Fabbri:2009ta}. In fact, matter fields  have degrees of freedom, corresponding to solutions of systems of differential equations, yielding the highest-order time derivative for the field. Constraints that obstruct the appearance of such highest-order time derivatives should, thus, be imposed \cite{Fabbri:2009ta}.
For higher-spin matter fields, inconsistencies can be circumvented by the Velo-Zwanziger procedure  \cite{VeloZwan1,VeloZwan2}. In Ref. \cite{VeloZwan2} a similar prescription is employed to introduce an appropriate Dirac operator. 
The Velo-Zwanziger problem has been  generalized to encompass torsion  \cite{zu1,zu2,zu3}
and the torsion-spin coupling  \cite{13}. 
The constraints imposed by the Velo-Zwanziger analysis are shown to be strengthened by the  backreaction due to the torsion, having also further constraints coming from 
the spin-torsion coupling. These constraints can avoid the existence of intricate matter fields, restricting the amount of fields available to be used in this kind of theory.

To simplify further, we assume $v\cdot\Gamma=v_{\mu}\gamma^{\mu}$.
Denoting the conventional free field Lagrangian for $\chi$ by 
\begin{equation}
\mathcal{L}_{0}=\frac{i}{2}\bar{\chi}\Gamma^{\nu}\stackrel{\leftrightarrow}{\partial_{\nu}}\chi-m\bar{\chi}\chi\thinspace,
\end{equation}
equation\,\eqref{eleclagran} can be written as, 
\begin{eqnarray}
{\mathcal{L}}={\mathcal{L}}_{0}+{\Re}v_{\mu}[i\bar{\chi}\stackrel{\leftrightarrow}{\partial^{\mu}}\chi-2m\bar{\chi}\gamma^{\mu}\chi]-i{\Im}v_{\mu}[\bar{\chi}\sigma^{\mu\nu}\stackrel{\leftrightarrow}{\partial_{\nu}}\chi].\label{aeredef}
\end{eqnarray}
Comparing with \eqref{gamaLV}, the simultaneous choice of LV parameters
${\mathfrak{e}^{\mu}}=2{\Re}v^{\mu}$ and $a^{\mu}=2m{\Re}v^{\mu}$
may be entirely attributed to the field redefinition, and therefore
does not introduce a real LV. In a different perspective, one might
say that in the LV theory defined by\,\eqref{eleclagran}, either
${\mathfrak{e}^{\mu}}$ or $a^{\mu}$ can be eliminated
via a field redefinition. The term ${\Im}v_{\mu}$ indicates that
the choice $2{\Im}\,v_{[\mu}g_{\rho]\nu}=g_{\rho\mu\nu}$ eliminates
the terms proportional to $g_{\rho\mu\nu}$ in\,\eqref{eleclagran}.
The parameters $m_{5}$, $a_{\mu}$, ${\mathfrak{e}_{\mu}}$,
$f_{\mu}$, and $c_{[{\mu\nu}]}$ in\,\eqref{eleclagran} can also
be removed\,\cite{colladay2}.

Now we review how this discussion can be extended to the Riemann-Cartan
torsional context. By using latin indexes to label local Lorentz coordinates
and greek indexes for spacetime ones, the Minkowski metric is related
to the curved-spacetime metric $g_{\mu\nu}$ via the \emph{vierbein}
$\vb\mu a$, by the relation $g_{\mu\nu}=\vb\mu a\vb\nu b\et_{ab}$.
{The determinant of the }{\emph{vierbein}}{{}
is denoted by $e$ and the charge of the electron is denoted by $-q$}.
For the spacetime covariant derivative, the connection is assumed
to be metric compatible. In addition, curved-spacetime indexes are
corrected with the Cartan connection $\Ga_{\pt\la\mu\nu}^{\la}$,
namely, 
\begin{equation}
\nabla_{\mu}\vb\nu a=\prt_{\mu}\vb\nu a+\lulsc\mu ab\vb\nu b-\Ga_{\pt\al\mu\nu}^{\al}\vb\al a\thinspace.
\end{equation}
The contortion tensor is defined as 
\begin{equation}
K_{\mu\nu}^{\lambda}=\frac{1}{2}(T_{\mu\nu}^{\,\,\,\lambda}-T_{\nu\,\,\mu}^{\,\lambda}-T_{\,\,\,\mu\nu}^{\lambda})\thinspace,
\end{equation}
and the curvature as $R_{\pt\ka\la\mu\nu}^{\ka}=\mathring{R}_{\pt\ka\la\mu\nu}^{\ka}+\nabla_{[\mu}K_{\pt{\ka}\nu]\la}^{\ka}+K_{\pt\al[\mu\nu]}^{\al}K_{\pt{\ka}\al\la}^{\ka}+K_{\pt\al[\mu\la}^{\al}K_{\pt{\ka}\nu]\al}^{\ka}$,
where $\mathring{R}_{\pt\ka\la\mu\nu}^{\ka}$ denotes the usual Riemann
curvature tensor in the absence of torsion. The source of contortion
may be considered as a Kalb-Ramond field $B_{\alpha\beta}$, through
$K_{\;\,\alpha\beta}^{\rho}=-\frac{1}{\kappa^{3/2}}H_{\;\,\alpha\beta}^{\rho}$,
where $H_{\rho\alpha\beta}=\partial_{[\rho}B_{\alpha\beta]}$ and
$\kappa$ denotes the coupling constant.
However, our discussion will not depend on this identification, being
a generic geometric contortion considered hereon.

Signatures of torsion, in the context of both minimal and non-minimal
couplings to fermions, are phenomenologically and experimentally abundant\,\cite{Kostelecky:2007kx}.
The minimal coupling between torsion and SM fields is realized through
covariant derivatives. Nevertheless, non-minimal couplings are also
an option. In fact, we shall prove that when flagpoles spinors are
regarded, non-minimal couplings are the only possibility.

The essential variables here are the \emph{vierbein} and the spin
connection, since other variables such as curvature and torsion can
be expressed in terms of these. For example, the Cartan connection
reads $\Ga_{\pt\la\mu\nu}^{\la}=\uvb\la a(\prt_{\mu}\lvb\nu a-\lulsc\mu ba\lvb\nu b),$
whereas the torsion is given by $T_{\la\mu\nu}=\vb\la a(\prt_{[\mu}e_{\nu]a}+\omega_{[\mu|ab|}e_{\nu]}^{b})$.
Moreover, the spin connection is related to the \emph{vierbein} by
\begin{eqnarray}
\nsc\mu ab & = & \half e^{\nu[a}\prt_{[\mu}e_{\nu]}^{\;b]}-\half\uvb\al a\uvb\be b\vb\mu c\prt_{[\al}e_{\be]c}+K_{\nu\mu\la}\uvb\nu a\uvb\la b.
\end{eqnarray}
Hereon weak gravitational fields, $g_{\mu\nu}=\et_{\mu\nu}+h_{\mu\nu},$
are regarded, where $h_{\mu\nu}$ is a fluctuation. At leading order,
the \emph{vierbein} and spin connection can be expressed in terms
of small quantities, 
\begin{eqnarray}
\lvb\mu a & = & \et_{\mu a}+\ep_{\mu a}\approx\et_{\mu a}+\half h_{\mu a}+\ch_{\mu a},\qquad\quad e\approx1+\half h,\\
\lsc\mu ab & \approx & -\half\prt_{a}h_{\mu b}+\half\prt_{b}h_{\mu a}+\prt_{\mu}\ch_{ab}+K_{a\mu b}.\label{123321}
\end{eqnarray}

The basic non-gravitational fields for the Lorentz- and CPT-violating
QED extension in Riemann-Cartan spacetime are a fermion field $\ps$
and the photon field $A_{\mu}$. The action for the theory can be
expressed as a sum of partial actions for the fermion, for the photon
and for gravity. The fermion part of the action contains terms that
are dominant at low energies, involving fermions and their minimal
couplings to photons and gravity. In general, we also have to consider
higher order terms involving fermions and photons that are non-renormalizable,
non-minimal and higher order in the gravitational couplings, as well
as field operators of dimension greater than four that couple curvature
and torsion to the matter and photon fields.

The fermion part of the action (\ref{eleclagran}) for the QED extension
can be written as {
\begin{equation}
S=\int d^{4}x\,e\left(\half i\ivb\mu a\bar{\ps}\Ga^{a}\lrDmu\ps-\bar{\ps}M\ps\right),\label{qedxps}
\end{equation}
}where the usual $U(1)$ covariant derivative is given by $\nabla_{\mu}\ps\equiv\prt_{\mu}\ps+\frac{1}{4}i\nsc\mu ab\si_{ab}\ps-iqA_{\mu}\ps$\,\cite{Fabbri:2010pk}.
In addition, Eqs.\,(\ref{gamaLV}) and\,(\ref{mmm}) read, respectively,
in terms of multivector structure of the \emph{vierbein}, 
\begin{eqnarray}
\Ga^{a} & = & \ga^{a}-c_{\mu\nu}\uvb\nu a\ivb\mu b\ga^{b}+d_{\mu\nu}\ivb\mu b\uvb\nu a\ga^{b}\ga_{5}-{\mathfrak{e}_{\rho}}\uvb\rho a-if_{\mu}\ga_{5}\uvb\mu a-\half g_{\la\mu\nu}\ivb\la b\uvb\nu a{\mathfrak{e}^{\mu}}c\si^{bc}\thinspace,\label{gamdef}\\
M & = & m+a_{\mu}\ivb\mu a\ga^{a}+\half H_{\mu\nu}\ivb\mu a\ivb\nu b\si^{ab}+b_{\mu}\ivb\mu a\ga_{5}\ga^{a}+im_{5}\ga_{5}.\label{mdef}
\end{eqnarray}
Hence the Dirac equation in Riemann-Cartan spacetimes reads \cite{colladay2}
\begin{eqnarray}
 &  & i\ivb\mu b\Ga^{b}\nabla_{\mu}\ps+\half\ivb\mu a\nsc\mu bc(i\et_{\pt{a}b}^{a}\Ga_{c}-\frac{1}{4}[\si_{bc},\Ga^{a}])\ps-\half i\tor\rho\rho\mu\ivb\mu a\Ga^{a}\ps-M\ps=0.\label{direq}
\end{eqnarray}
The LV terms involving $M$ contribute to the Dirac equation in a
minimal way, as assumed in non-derivative couplings. Nevertheless,
those terms involving $\Ga^{a}$ emerge both minimally and through
commutation with the Lorentz generators in the covariant derivative.
In particular, the Lorentz-invariant parts of the terms in Eq.\ \eqref{direq}
cancel out.

Lorentz violating corrections can be then explored in the context
of singular spinor fields. In fact, the QED extension in Minkowski
and Riemann-Cartan spacetimes differ by weak gravitational couplings.
In this regime, given by Eq.\,(\ref{123321}), the Lagrangian terms
that are linear read 
\begin{equation}
\cl_{\ps}\sim-i(c_{{\rm effective}})_{\mu\nu}\bar{\ps}\ga^{\mu}\prt^{\nu}\ps-(b_{{\rm effective}})_{\mu}{\overbrace{\bar{\ps}\ga_{5}\ga^{\mu}\ps}^{=\,K^{\mu}}},
\end{equation}
where 
\begin{eqnarray}
(c_{{\rm effective}})_{\mu\nu} & \equiv & c_{\mu\nu}+\ch_{\mu\nu}-\half h_{\mu\nu}\,,\qquad(b_{{\rm effective}})_{\mu}\equiv b_{\mu}+\frac{1}{8}\ep_{\mu\nu\rho\sigma}T^{\nu\rho\sigma}-\frac{1}{4}\prt^{\nu}\ch^{\rho\sigma}\ep_{\mu\nu\rho\sigma}.\label{effcoeff}
\end{eqnarray}
In this expression, leading-order terms arising from the scaling of
the \emph{vierbein} determinant $e$ are neglected, for being LI.
Equations\,\eqref{effcoeff} show that, at leading order, a weak
background metric is governed by $c_{\mu\nu}$, whereas the torsion
is effectively ruled by a $b_{\mu}$ term\,\cite{shapiro}. The latter
is a CPT-violating term, so the presence of background torsion can
mimic CPT violation. If the Lagrangian\,(\ref{effcoeff}) models
a flagpole spinor field, that satisfies {$K^{\mu}=\bar{\ps}\ga_{5}\ga^{\mu}\ps=0$},
the $b_{{\rm effective}}$ term is irrelevant in such a model. Hence,
flagpole fermions are not sensitive to this type of LV.

Considering the mapping between LI and LV spinor fields given by
Eq.\,(\ref{spintrans}), they can be used to show that, at leading
order in coefficients for Lorentz violation, there are no physical
effects from the coefficients {$\mathfrak{e}_{\mu}$},
\f, or from the antisymmetric parts of $c_{\mu\nu}$.

When non-minimal gravitational couplings are taken into account, operators
of mass dimension four or less can be analyzed. In the QED extension,
such non-minimal operators are not leading couplings, and the only
gauge-invariant ones are products of the torsion with fermion bilinear
covariants. The Lorentz invariant possibilities are {
\begin{eqnarray}
\cl_{{\rm LI}}\! & = & \!(aT_{\pt{\la}\la\mu}^{\la}+a_{5}\ep_{\mu\nu\rho}\,T^{\mu\nu\rho})\overbrace{\bar{\ps}\ga^{\mu}\ps}^{=\,J^{\mu}}+(bT_{\pt{\la}\la\sigma}^{\la}+b_{5}T^{\mu\nu\rho}\ep_{\mu\nu\rho\sigma})\overbrace{\bar{\ps}\ga_{5}\ga^{\mu}\ps}^{=\,K^{\sigma}}.\label{LLI}
\end{eqnarray}
} The $b_{5}$ coupling is minimal, whereas the other ones are non
minimal. The LV possibilities are {
\begin{eqnarray}
\cl_{{\rm LV}} & \!=\! & k_{\mu\nu\rho}T^{\mu\nu\rho}\overbrace{\bar{\ps}\ps}^{=\,\sigma}\!+\!k_{\mu\nu\rho\sigma}T^{\mu\nu\rho}\overbrace{\bar{\ps}\ga^{\sigma}\ps}^{=\,J^{\sigma}}+k_{\mu\nu\rho\sigma\tau}T^{\mu\nu\rho}\overbrace{\bar{\ps}\si^{\sigma\tau}\ps}^{\stackrel{=\,-2iS^{\sigma\tau}}{}}\!+\!k_{5\mu\nu\rho\sigma}T^{\mu\nu\rho}\overbrace{\bar{\ps}\ga_{5}\ga^{\sigma}\ps}^{K^{\sigma}}\!\nonumber \\
 &  & \qquad+k_{5\mu\nu\rho}T^{\mu\nu\rho}\overbrace{\bar{\ps}\ga_{5}\ps}^{\omega}.\label{LLV}
\end{eqnarray}
The $k_{\mu\gamma\alpha\beta}$ must have the symmetries of the Riemann
tensor \cite{colladay4}. In our case, we use the equation (33) of
\cite{kostlecky4}, with the following form $k_{\mu\gamma\alpha\beta}=\frac{1}{2}(\eta_{\mu\alpha}h_{\gamma\beta}-\eta_{\gamma\alpha}h_{\mu\beta}+\eta_{\gamma\beta}h_{\mu\alpha}-\eta_{\mu\beta}h_{\gamma\alpha})$,
where $h_{\mu\nu}$ is the weak-field background metric.} If Lorentz
violation is suppressed, and the torsion is also small, then all terms
in Eq.\,\eqref{LLV} are subdominant. Nonetheless, all the above
operators may be of interest in more exotic scenarios. For example,
the presence of a Higgs doublet in the SME allows for other types
of non-minimal gravitational couplings of dimension four or more,
including ones involving both curvature and torsion. Operators of
dimension greater than four generically come with Planck-scale suppression\,\cite{Kostelecky:2007kx}.
Therefore, effects of dimension five Lorentz invariant operators suppressed
by the inverse of the Planck mass $m_{P}$ are comparable, in magnitude,
to those of a dimension four operator involving a coefficient for
LV suppressed by $m_{P}$, for example.

It is worth to emphasize that, when flagpole spinor fields are taken
into account, the second and fourth terms on the right-hand side of
Eq.\,(\ref{LLV}) identically vanish, as well as the second and fourth
terms on the right-hand side of Eq.\,(\ref{LLI}). Hence, flagpole
fermions have a restricted range of couplings, when compared to Dirac
fermions.

By taking into account the Lounesto classification, we already know
that in Riemann-Cartan geometry in a $f(R)$ conformal gravity setup,
flag-dipole type-(4) spinor fields are solutions of the Dirac equation\,\cite{riemanncartan}.
In this context, flagpole also play a prominent role when torsion
is taken into account, as the last term on the right hand side in
Eq.\,(\ref{LLI}) is zero, as well as the first, the second, and
the last term on the right hand side in Eq.\,(\ref{LLV}). Hence,
flagpole spinor fields are examples of singular spinor fields which
we have proved to be less sensitive to Lorentz violation.

Fermionic matter fields in Riemann-Cartan spacetimes can be governed
by a Lagrangian with arbitrary torsion couplings. In a constant torsion
approximation setup, the torsion couplings can be replaced by background
solutions to the torsion field equations. The corresponding effective
Lagrange density reads\,\cite{Kostelecky:2007kx} 
\begin{eqnarray}
\cl & \sim & \half i\bar{\ps}\ga^{\mu}\stackrel{\leftrightarrow}{\partial_{\mu}}\ps-m\bar{\ps}\ps+\cl_{{\rm LI(4)}}+\cl_{{\rm LI(5)}}\thinspace,
\end{eqnarray}
where all possible independent constant-torsion couplings of mass
dimensions four and five are respectively given by {
\begin{align}
\cl_{{\rm LI(4)}} & =(a_{1}T_{\;\rho\mu}^{\rho}+a_{3}\mathfrak{A}_{\mu}){\overbrace{\bar{\ps}\ga^{\mu}\ps}^{=\,J^{\mu}}}+(a_{2}T_{\;\rho\mu}^{\rho}+a_{4}\mathfrak{A}_{\mu}){\overbrace{\bar{\ps}\ga_{5}\ga^{\mu}\ps}^{=\,K^{\mu}}}+\half i\mathring{a}_{1}T^{\mu}\bar{\ps}\stackrel{\leftrightarrow}{\partial_{\mu}}\ps\label{lag1}\\
\cl_{{\rm LI(5)}} & =\half\left(\mathring{a}_{2}T^{\mu}+\mathring{a}_{4}\mathfrak{A}^{\mu}\right)\bar{\ps}\ga_{5}\stackrel{\leftrightarrow}{\partial_{\mu}}\ps+\half i\mathring{a}_{3}\mathfrak{A}^{\mu}\bar{\ps}\stackrel{\leftrightarrow}{\partial_{\mu}}\ps+\half i\left(\mathring{a}_{5}{M^{\nu}}_{\mu\lambda}+\mathring{a}_{6}T_{\;\rho\mu}^{\rho}+\mathring{a}_{7}\mathfrak{A}_{\mu}\right)\bar{\ps}\stackrel{\leftrightarrow}{\partial_{\nu}}\si^{\mn}\ps\nonumber \\
 & +\half i\left(\mathring{a}_{8}\cep^{\la\ka\mn}T_{\rho\lambda}^{\rho}+\mathring{a}_{9}\cep^{\la\ka\mn}\mathfrak{A}_{\la}\right)\bar{\ps}\stackrel{\leftrightarrow}{\partial_{\ka}}\si_{\mn}\ps,\label{lag}
\end{align}
}where the $a^{A}$ {[}$\mathring{a}^{A}${]} denote mass dimension
four {[}five{]} coupling constants, 
\begin{equation}
M_{\alpha\mu\nu}=\frac{1}{3}\left(T_{\alpha\mu\nu}+T_{\mu\alpha\nu}+T_{\mu}g_{\alpha\mu}\right)-\frac{1}{3}(\mu\leftrightarrow\nu)\thinspace,
\end{equation}
and $\mathfrak{A}^{\mu}=\frac{1}{6}\epsilon^{\alpha\beta\gamma\mu}T_{\alpha\beta\gamma}$
\cite{Kostelecky:2007kx,kostlecky4}.

It is worth to emphasize that the Lagrangian in Eq. (\ref{lag1})
is gauge invariant, whereas Eq. (\ref{lag}) is not gauge invariant.
The minimal coupling is obtained in the particular case where $a_{4}=3/4$
and all other couplings vanish. By considering Eqs.\,(\ref{Elko11})
to\,(\ref{dirac21}), we can see that type-(5), flagpole spinor fields,
are described by 
\begin{eqnarray}
\cl_{(4)}^{{\rm type-(5)}} & \sim & {(a_{1}T_{\;\rho\mu}^{\rho}+ a_{3}\mathfrak{A}_{\mu})\overbrace{\bar{\ps}\ga^{\mu}\ps}^{=\,J^{\mu}}}\thinspace,\label{lag}
\end{eqnarray}
and at least half of the couplings between flagpole fermions, associated
with dimension four coupling constants, and torsion, are vanishing.
We have shown that, in Riemann-Cartan spacetimes, flagpoles spinors
are less sensitive to Lorentz violation. Recent experimental searches
for Lorentz violation are exploited to extract new constraints involving
independent torsion components down to levels of order $10^{-31}$
GeV\,\cite{Kostelecky:2007kx}. Although exceptional sensitivity
to spacetime torsion can be achieved by searching for its couplings
to Dirac fermions, flagpole spinors are shown to be less sensitive
to torsion.

\section{\label{sec:A-bridge-between}A bridge between Lorentz symmetry and
Lorentz symmetry violation}

Once we proved that singular spinors are less sensitive to torsion
couplings in LV scenarios, we now discuss the relation between bilinear
covariant in the apparently LV framework and the standard covariant
bilinears. Since bilinear covariants realize the observables in theories
involving fermionic fields, this is a question of upmost physical
relevance. We will discuss how a general LV spinor field can be transliterated
into the Lorentz-invariant ones, and show that this relation mixes
the spinor classes as defined in the Lounesto classification.

One formal concern when considering LV is what part of the Lorentz
symmetry is broken and what part does remain. Hence, if LV is obtained
by allowing a larger class of transformations that are, in general,
non-linear and dependent on coordinates and derivatives, this formalism
should embed the Lorentz group into a larger group. This might make
it necessary to describe how the Lorentz group places inside this
larger symmetry group. However, by focusing on the study of spinors
via the bilinear invariants, we do not need to address these questions.
In fact, bilinear covariants of any spinor does concern neither to
the representation with respect to the residual symmetry nor to the
group associated with the residual symmetry. This is due to the fact
that any symmetry group can be embedded in some Spin group that, up
to dimension five, can be defined by the invertible elements $R$
of the twisted Clifford-Lipschitz group that satisfy $R^{\dagger}R=I$\,\cite{lou2,oxford}.
Hence, any residual symmetry shall not be apparent when bilinear covariants
are taken into account. When spinors are taken into account, then
obviously the content of the residual symmetry is important. Nevertheless,
our aim to take into account the observables, namely the bilinear
covariants, make the group content to be shortcut. It is worth to
mention that by taking a classical spinor $\xi$ which satisfies $\xi^{\dagger}\gamma_{0}\psi\neq0$,
the original spinor $\psi$ can be recovered from its aggregate $\mathbf{Z}$,
which is given by 
\begin{eqnarray}
\mathbf{Z}=\sigma+\mathbf{J}+i\mathbf{S}+i\mathbf{K}\gamma_{0123}+\omega\gamma_{0123}\,,\label{Z}
\end{eqnarray}
using the Takahashi algorithm\,\cite{oxford}.

Besides removing spurious Lorentz violation terms from the Lagrangian,
we want to investigate the effect of these field redefinitions in
the Lounesto classification for the transformed spinors. We accomplish
this by relating the observables in the Lorentz violating framework
with the bilinear covariants in the standard LI theory. Hence the
regular (in particular Dirac) spinors and the singular spinors (encompassing
Weyl, Majorana and Elko spinors, among others) can have a dual description
in terms of a Lorentz violating structure.

We start by listing the covariant bilinears in the Lorentz violating framework,
\begin{eqnarray}
\sigma_{\chi} & = & \bar{\chi}\chi,\nonumber \\
\mathbf{J}_{\chi} & = & \bar{\chi}\gamma_{\mu}\chi{\rm e}^{\mu},\nonumber \\
\mathbf{S}_{\chi} & = & \frac{1}{2}\bar{\chi}i\gamma_{\mu\nu}\chi{\rm e}^{\mu}\wedge{\rm e}^{\nu},\nonumber \\
\mathbf{K}_{\chi} & = & \bar{\chi}\gamma_{5}\gamma_{\mu}\chi{\rm e}^{\mu},\nonumber \\
\omega_{\chi} & = & \bar{\chi}\gamma_{5}\chi.
\end{eqnarray}
Next, we relate these invariants with the corresponding ones for the
transformed (LI) spinors. After some calculation, we find 
\begin{align}
\sigma_{\psi} & =\Delta\sigma_{\chi}+\Omega_{\sigma}\thinspace,\label{eq:upsigma}
\end{align}
where {$\Delta=1+2\Im{\rm \theta}+\vert{\rm \theta}\vert^{2}+2\tilde{C}_{\mu}x^{\mu}\Re{\rm \theta}+\tilde{C}_{\mu}\tilde{C}_{\nu}x^{\mu}x^{\nu}$},
and the explicit form of $\Omega_{\sigma}$ is given in the Appendix.
The first terms of this expression read 
\begin{align}
\Omega_{\sigma} & =\chi^{\dagger}\gamma_{0}v\cdot\Gamma\chi+B_{\mu}\chi^{\dagger}\gamma_{0}\partial^{\mu}\chi+\tilde{B}_{\mu}\chi^{\dagger}\gamma_{0}\gamma_{5}\partial^{\mu}\chi+\cdots
\end{align}
For the remaining invariants we obtain similar expressions,\begin{subequations}\label{eq:identifications}
\begin{align}
J_{\alpha}^{\psi} & =\Delta J_{\alpha}^{\chi}\ +\Omega_{J_{\alpha}},\,(\gamma_{0}\mapsto\gamma_{0}\gamma_{\alpha})\\
S_{\alpha\beta}^{\psi} & =\Delta S_{\alpha\beta}^{\chi}+\Omega_{S_{\alpha\beta}},\,(\gamma_{0}\mapsto\gamma_{0}\gamma_{\alpha\beta})\\
K_{\alpha}^{\psi} & =\Delta K_{\alpha}^{\chi}\ +\Omega_{K_{\alpha}},\,(\gamma_{0}\mapsto\gamma_{0}\gamma_{5}\gamma_{\alpha})\\
\omega^{\psi} & =\Delta\omega^{\chi}\ +\Omega_{\omega},\,(\gamma_{0}\mapsto\gamma_{0}\gamma_{5})
\end{align}
\end{subequations}where the identifications in the parentheses mean
that, for example, $\Omega_{J_{\alpha}}$ is obtained from $\Omega_{\sigma}$
by the substitution $\gamma_{0}\mapsto\gamma_{0}\gamma_{\alpha}$,
and similarly for the remaining invariants.

Using these expressions, we can apply the Lounesto classification
for the transformed (LI) spinors. Consider for example the case $\sigma_{\psi}\neq0,\;\omega_{\psi}\neq0$,
corresponding to a type-1 spinor in the LI theory. Since $\sigma_{\psi}=\Delta\sigma_{\chi}+\Omega_{\sigma}$
and $\omega_{\psi}=\Delta\omega_{\chi}\ +\Omega_{\omega}$, depending
on the values of $\Omega_{\sigma}$ and $\Omega_{\omega}$, one can
have either $\sigma_{\chi}$ and $\sigma_{\omega}$ equal to zero
or not, and therefore the field redefinition may relate a Lorentz
invariant type-1 spinor to several types of Lorentz violating spinors,
such as $\sigma_{\chi}=0=\omega_{\chi}\ $ (types 4, 5, and 6), and
others. A comprehensive list of possibilities is given below, where
we shall assume that the functions $\Delta$ and $\Omega$ are non-vanishing: 
\begin{itemize}
\item[$1_{\psi}$)] $\sigma_{\psi}\neq0,\;\;\;\omega_{\psi}\neq0$.

Since $\sigma_{\psi}\neq0$ and $\sigma_{\psi}=\Delta\sigma_{\chi}+\Omega_{\sigma}$,
we list below all possibilities, depending upon whether $\sigma_{\chi}$
either does or does not equal zero, as well as $\omega_{\chi}$: 
\begin{enumerate}
\item[$i)$] $\sigma_{\chi}=0=\omega_{\chi}\ $. These conditions are correspondent
to the type-(4), type-(5), and type-(6) spinor fields \textemdash{}
respectively flag-dipoles, flagpoles, and dipoles. 
\item[$ii)$] $\sigma_{\chi}=0$ and $\omega_{\chi}\neq0$, being compatible to
type-(3) regular spinor fields. The condition $\sigma_{\chi}=0$ is
consistent with $\sigma_{\psi}\neq0$. 
\item[$iii)$] $\sigma_{\chi}\neq0$ and $\omega_{\chi}=0$. This case regards type-(2)
regular spinor fields. The condition $\omega_{\chi}=0$. is consistent
with $\omega_{\sigma}\neq0$ 
\item[$iv)$] $\sigma_{\chi}\neq0$ and $\omega_{\chi}\neq0$, corresponding to
type-(1) Dirac spinor fields. 
\end{enumerate}
\item[$2_{\psi}$)] $\sigma_{\psi}\neq0,\;\;\;\omega_{\psi}=0$.\label{dirac1b}

Although the condition $\sigma_{\psi}\neq0$ is consistent with both
the possibilities $\sigma_{\chi}=0$ and $\sigma_{\chi}\neq0$ (clearly
the condition $\sigma_{\chi}\neq0$ is consistent with $\sigma_{\psi}\neq0$
if $\Delta\sigma_{\chi}\neq-\Omega_{\sigma}$), the condition $\omega_{\psi}=0$
yields $\Delta\omega_{\chi}=-\Omega_{\sigma}$, which does not vanish. 
\begin{enumerate}
\item[$i)$] $\sigma_{\chi}=0$ and $\omega_{\chi}\neq0$. This case corresponds
to the type-(3) regular spinor fields. The condition $\sigma_{\chi}=0$
is consistent with $\sigma_{\psi}\neq0$, however since $\omega_{\chi}\neq0$,
the additional condition $\Delta\omega_{\chi}=-\Omega_{\sigma}$ must
be imposed. 
\item[$ii)$] $\sigma_{\chi}\neq0$ and $\omega_{\chi}\neq0$. This case regards
type-(1) Dirac spinor fields. 
\end{enumerate}
\item[$3_{\psi}$)] $\sigma_{\psi}=0,\;\;\;\omega_{\psi}\neq0$.\label{dirac2b} The
condition $\omega_{\psi}\neq0$ is consistent with both complementary
$\omega_{\chi}=0$ and $\omega_{\chi}\neq0$

To summarize: 
\begin{enumerate}
\item[$i)$] $\omega_{\chi}=0$ and $\sigma_{\chi}\neq0$, corresponding to type-(2)
regular spinor fields. The condition $\omega_{\chi}=0$ is consistent
with $\omega_{\psi}\neq0$, however since $\sigma_{\chi}\neq0$, the
additional condition $\sigma_{\psi}=\Delta\sigma_{\chi}+\Omega_{\sigma}\neq0$
must be imposed. 
\item[$ii)$] $\sigma_{\chi}\neq0$ and $\omega_{\chi}\neq0$. This case regards
type-(1) Dirac spinor fields. 
\end{enumerate}
\item[$4_{\psi}$)] $\sigma_{\psi}=0=\omega_{\psi},\;\;\;\mathbf{K}_{\psi}\neq0,\;\;\;\mathbf{S}_{\psi}\neq0$.\label{tipo4b} 
\item[$5_{\psi}$)] $\sigma_{\psi}=0=\omega_{\psi},\;\;\;\mathbf{K}_{\psi}=0,\;\;\;\mathbf{S}_{\psi}\neq0$.\label{type-(5)1b} 
\item[$6_{\psi}$)] $\sigma_{\psi}=0=\omega_{\psi},\;\;\;\mathbf{K}_{\psi}\neq0,\;\;\;\mathbf{S}_{\psi}=0$. 
\end{itemize}
\noindent All singular spinor fields $4_{\psi}$), $5_{\psi}$), and
$6_{\psi}$) are defined by the condition $\sigma_{\psi}=0=\omega_{\psi}$,
which implies that $\Delta\sigma_{\chi}=-\Omega_{\sigma}(\neq0)$
and that $\Delta\omega_{\chi}=-\Omega_{\omega}(\neq0)$. Therefore,
singular spinors in the Lorentz violating theory are always related
to a regular spinor in the corresponding Lorentz invariant model.
This discussion can be summarized in Table\,\ref{table}, that describes
the possibility of mapping observables in the Lorentz violating framework
into observables in the Lorentz covariant models.

\begin{table}
\centering{}%
\begin{tabular}{||r|r||r|r||}
\hline 
 & LV Spinor Fields  & Covariant Spinor Fields  & \tabularnewline
\hline 
\hline 
class ($1_{\Psi}$)  & $\Psi$-regular  & Regular\qquad{}  & class (1)\tabularnewline
 &  & Regular\qquad{}  & class (2)\tabularnewline
 &  & Regular\qquad{}  & class (3)\tabularnewline
 &  & Flag-dipoles\quad{}  & class (4)\tabularnewline
 &  & Flagpoles\quad{}  & class (5)\tabularnewline
 &  & Dipoles\qquad{}  & class (6)\tabularnewline
\hline 
class ($2_{\Psi}$)  & $\Psi$-regular  & Regular\qquad{}  & class (3)\tabularnewline
 &  & Regular\qquad{}  & class (1)\tabularnewline
\hline 
class ($3_{\Psi}$)  & $\Psi$-regular  & Regular\qquad{}  & class (2)\tabularnewline
 &  & Regular\qquad{}  & class (1)\tabularnewline
\hline 
class ($4_{\Psi}$)  & $\Psi$-flag-dipole  & Regular\qquad{}  & class (1)\tabularnewline
\hline 
class ($5_{\Psi}$)  & $\Psi$-flagpole  & Regular\qquad{}  & class (1)\tabularnewline
\hline 
class ($6_{\Psi}$)  & $\Psi$-dipole  & Regular\qquad{}  & class (1)\tabularnewline
\hline 
\end{tabular}\caption{\label{table}Correspondence among LV spinor fields ($\chi$) and
the LI $\Psi$-spinor fields under Lounesto spinor field classification.}
\end{table}

The possibility of mapping between different spinor classes has already
been pointed out in the literature. In fact, it is known that a regular
spinor field can be mapped into any covariant spinor field, including
exotic mappings \cite{Bernardini:2012sc}. Such mappings are
formally consistent and were accomplished both in the kinematical
and the dynamical contexts. Nevertheless, all these studies were developed
under a Lorentz invariant framework, that is generalized in this present
work to include the possibility of LV, thus enlarging considerably
the class of models that can be related by these transformations.

As an example of the kind of relation discussed here, let us construct
a transformation that maps a Dirac spinor $\chi$ in the LV framework
to a singular spinor $\psi$ (Majorana, flagpole, type) in the LI
one. Without loss of generality, in the Weyl representation these
spinors are eigenspinors of the charge conjugation operator and can
be parametrized as follows, 
\begin{equation}
\chi=\begin{pmatrix}a_{0}\\
a_{1}\\
a_{2}\\
a_{3}
\end{pmatrix}=\begin{pmatrix}\chi_{1}\\
\chi_{2}
\end{pmatrix}\in\mathbb{C}^{4},\quad\qquad\psi=\begin{pmatrix}-i\beta^{*}\\
i\alpha^{*}\\
\alpha\\
\beta
\end{pmatrix}=\begin{pmatrix}\psi_{1}\\
\psi_{2}
\end{pmatrix}\in\mathbb{C}^{4}\,,\label{eq:example1}
\end{equation}
and we can further choose $a_{0}=a_{2}=0$\,\cite{Cavalcanti:2014wia}.
The necessary transformation is of the form $\psi=(1+v_{\mu}\gamma_{5}\gamma^{\mu})\chi$,
which is equivalent to $\psi-\chi=v_{\mu}\gamma_{5}\gamma^{\mu}\chi$.
Defining 
\begin{align}
T & \equiv v_{\mu}\gamma_{5}\gamma^{\mu}=\begin{pmatrix}0 & -v_{\mu}\sigma^{\mu}\\
v_{0}I_{2}-v_{k}\sigma^{k} & 0
\end{pmatrix}\thinspace,
\end{align}
{where $\sigma^{0}=I_{2}$, and $\sigma^{k}$ are the
Pauli matrices, namely, }$T\chi=\binom{-v_{\mu}\sigma^{\mu}(\chi_{2})}{v_{0}I-v_{k}\sigma^{k}(\chi_{1})}=\binom{\psi_{1}-\chi_{1}}{\psi_{2}-\chi_{2}}$.
Hence we can obtain consistency conditions between the components
of the spinors, namely, $i\beta^{*}/a_{3}=-\al/a_{1}\equiv\delta\in\mathbb{C}\thinspace,$
together with $v_{1}-iv_{2}=\delta$, which characterizes the desired
transformations. The end result is that, in the classical level, the
LV model of a Dirac spinor $\chi$ governed by the Lagrangian 
\begin{equation}
\mathcal{L}_{0}=\frac{i}{2}\bar{\chi}\gamma^{\nu}\stackrel{\leftrightarrow}{\partial_{\nu}}\chi-m\bar{\chi}\chi-iv_{\mu}\bar{\chi}\gamma_{5}\stackrel{\leftrightarrow}{\partial^{\mu}}\chi
\end{equation}
is actually physically identical to the standard LI model of a free
Majorana spinor. One may recall that regular spinors in the class 1 of the Lounesto's classification have eight degrees of freedom, whereas Majorana spinors have four degrees of freedom.
Equation\,(\ref{spintrans}) is a constraint that reduces two degrees
of freedom of the Dirac spinor. Moreover the lacking degree of freedom
is counted when the $U(1)$ gauge is taken into account for the Dirac
spinor, but not for the Majorana one\,\cite{lou2}.

A more natural mapping between spinors, alternatively to Eq.\,(\ref{spintrans})
would be the one leading Weyl spinors, that are a particular class
of dipole spinors with $U(1)$ gauge symmetry, to Dirac regular spinors,
what moreover enables Penrose flags to be attached to Weyl spinors.
However the relationship between Dirac and Weyl spinors is widely
explored throughout the literature of QFT. Hence, we opt to obtain
a mapping between regular spinor fields and flag-dipole ones, since
flag-dipole spinors are important solutions of the Dirac equation
in $f(R)$, torsional, gravity\,\cite{riemanncartan}. In fact, consider
regular and type 4 spinors respectively\,\cite{Cavalcanti:2014wia},
{
\begin{align}
\chi= & \begin{pmatrix}a_{0}\\
a_{1}\\
a_{2}\\
a_{3}
\end{pmatrix}\in\mathbb{C}^{4}\,,\text{ with }\begin{matrix}a_{0}\neq-\frac{a_{1}a_{2}a_{3}^{*}}{||a_{2}||^{2}}\end{matrix}\\
\psi= & \begin{pmatrix}c_{0}\\
c_{1}\\
c_{2}\\
c_{3}
\end{pmatrix}=\begin{pmatrix}-\frac{c_{1}c_{2}c_{3}^{*}}{||c_{2}||^{2}}\\
c_{1}\\
c_{2}\\
c_{3}
\end{pmatrix}\in\mathbb{C}^{4}\,,\text{ with }\begin{matrix}||c_{1}||^{2}\neq||c_{3}||^{2}\end{matrix}.
\end{align}
The transformation which relates the spinor fields follows by taking
 $(v\cdot\Gamma)\chi = \lambda \gamma^5\chi =  \psi - \chi$ 
in Eq.\,\eqref{spintrans}\,\cite{Cavalcanti:2014wia}. Then 
\begin{eqnarray}
 &  & \beta_{1}=1+\lambda=\frac{c_{2}}{a_{2}}=\frac{c_{3}}{a_{3}},\\
 &  & \beta_{2}=1-\lambda=\frac{c_{0}}{a_{0}}=\frac{c_{1}}{a_{1}},
\end{eqnarray}
and it yields $\lambda=\frac{\beta_{1}-\beta_{2}}{2}$.} In summary,
we have shown how a general class of field redefinitions can relate
between different classes of spinor fields in the LV and LI frameworks.
The physical equivalence between models here described holds in the
kinematical and dynamical classical aspects of these theories, leaving
open the interesting question of the quantum equivalence. Second quantization
of different classes of spinor fields can be accomplished, thus we
believe the investigation of the quantum aspects of the relations
exposed in this work is a viable and interesting topic to pursue.

\section{\label{sec:Concluding-remarks-and}Concluding remarks and outlook}

We considered the impact of the inclusion of Lorentz Violation in
theories involving general spinors in Riemann-Cartan spacetime. Couplings
between the spinor bilinear covariants and torsion have been proved
to provide a suitable signature to probe the sensitiveness of singular
spinor fields to LV parameters. Specifically, we proved that flagpole
spinor fields do not admit minimal coupling with torsion, and couplings
with mass dimension four coefficients are influenced by the choice
of flagpoles spinors, instead of the usual Dirac spinors. We also
proved that when flagpoles spinors are regarded, non-minimal couplings
between torsion and SM fields are the only feasible option. We conclude
that flagpole spinor fields are less sensitive to Lorentz violation.

The mapping between different classes of spinors in LV and Lorentz
covariant frameworks was also discusses, being a relevant computational
tool in QFT in LV setup. Explicit examples on how to map Dirac spinor
fields, in LV theories, into flagpole and flag-dipole spinors in Lorentz
invariant theories, were explicitly worked out.

The consideration of general spinor fields, as classified in the Lounesto
scheme, can lead to new dimensions in the exploration of new physics
involving Lorentz violation, and torsional gravity. \cite{Gomes:2008an}.
As an example, since the eigenspinors of the charge conjugation operator
with dual helicity\,\cite{cyleeI} are also examples of flagpoles
spinors, with the same structure as the flagpole in Eq.\,(\ref{eq:example1}),
phenomenological aspects in\,\cite{Dias:2010aa} can be further explored
in the context of LV scenarios\,\cite{Lee:2014opa}.

\textbf{\medskip{}
 }

\textbf{Acknowledgements.} This work was partially supported by Conselho
Nacional de Desenvolvimento Científico e Tecnológico (CNPq), Fundacão
de Amparo à Pesquisa do Estado de São Paulo (FAPESP), via the following
grants: CNPq 482874/2013-9, FAPESP 2013/22079-8 and 2014/24672-0 (AFF),
FAPESP 2015/10270-0, CNPq 303293/2015-2 (RR), and Universidade Federal
do ABC - UFABC PhD grant (JA).

\section{Appendix: The complete expression of $\Omega_{\sigma}$}

For the sake of completeness, we quote here the complete expression
for the function $\Omega_{\sigma}$ appearing in\,\eqref{eq:upsigma},

{
\begin{eqnarray*}
\Omega_{\sigma} & = & \Delta\{\tilde{B}_{\mu}\partial^{\mu}\chi^{\dagger}+i{\rm \theta}\tilde{B}_{\mu}\partial^{\mu}\chi^{\dagger}+i\tilde{B}_{\mu}\tilde{C}_{\alpha}x^{\alpha}\partial^{\mu}\chi^{\dagger}\}\gamma_{5}\gamma_{0}\chi\\
 & + & \{\chi^{\dagger}+\chi^{\dagger}(v\cdot\Gamma)^{\dagger}-i\tilde{C}_{\mu}x^{\mu}\chi^{\dagger}+\Delta(B_{\mu}\partial^{\mu}\chi^{\dagger}+\tilde{B}_{\mu}\partial^{\mu}\chi^{\dagger}\gamma_{5}+C_{\mu\nu}x^{\mu}\partial^{\nu}\chi^{\dagger})-i{\rm \theta}^{*}\chi^{\dagger}\}\gamma_{0}v\cdot\Gamma\chi\\
 & + & \{\chi^{\dagger}(v\cdot\Gamma)^{\dagger}+\Delta(B_{\mu}\partial^{\mu}\chi^{\dagger}+C_{\mu\nu}x^{\mu}\partial^{\nu}\chi^{\dagger})+i{\rm \theta}\chi^{\dagger}(v\cdot\Gamma)^{\dagger}+i\tilde{C}_{\alpha}x^{\alpha}\chi^{\dagger}(v\cdot\Gamma)^{\dagger}\\
 & + & i{\rm \theta}\Delta(B_{\mu}\partial^{\mu}\chi^{\dagger}+i\tilde{C}_{\alpha}B_{\mu}x^{\alpha}\partial^{\mu}\chi^{\dagger}+i{\rm \theta}C_{\mu\nu}x^{\mu}\partial^{\nu}\chi^{\dagger}+iC_{\mu\nu}x^{\mu}\tilde{C}_{\alpha}x^{\alpha}\partial^{\nu}\chi^{\dagger})\}\gamma_{0}\chi\\
 & + & \{\tilde{B}_{\alpha}\chi^{\dagger}+\Delta(\tilde{B}_{\alpha}\chi^{\dagger}(v\cdot\Gamma)^{\dagger}-\tilde{C}_{\mu}x^{\mu}\tilde{B}_{\alpha}\chi^{\dagger})+B_{\mu}\tilde{B}_{\alpha}\partial^{\mu}\chi^{\dagger}-\tilde{B}_{\mu}B_{\alpha}\partial^{\mu}\chi^{\dagger}\\
 & + & \tilde{B}_{\mu}\tilde{B}_{\alpha}\partial^{\mu}\chi^{\dagger}\gamma_{5}-\Delta\tilde{B}_{\mu}x^{\beta}C_{\beta\alpha}\partial^{\mu}\chi^{\dagger}+\tilde{B}_{\alpha}C_{\mu\nu}x^{\mu}\partial^{\nu}\chi^{\dagger}-i{\rm \theta}^{*}\Delta\tilde{B}_{\alpha}\chi^{\dagger}\}\gamma_{0}\gamma_{5}\partial^{\alpha}\chi\\
 & + & \{B_{\alpha}\chi^{\dagger}+C_{\beta\alpha}x^{\beta}\chi^{\dagger}-i\tilde{C}_{\mu}\Delta C_{\beta\alpha}x^{\mu}x^{\beta}\chi^{\dagger}+B_{\mu}B_{\alpha}\partial^{\mu}\chi^{\dagger}+B_{\mu}C_{\beta\alpha}x^{\beta}\partial^{\mu}\chi^{\dagger}\\
 & + & B_{\alpha}C_{\mu\nu}x^{\mu}\partial^{\nu}\chi^{\dagger}+C_{\mu\nu}C_{\beta\alpha}x^{\mu}x^{\beta}\partial^{\nu}\chi^{\dagger}-i\Delta({\rm \theta}^{*}B_{\alpha}\chi^{\dagger}-i{\rm \theta}^{*}C_{\beta\alpha}x^{\beta}\chi^{\dagger}\\
 & + & B_{\alpha}\chi^{\dagger}(v\cdot\Gamma)^{\dagger}+C_{\beta\alpha}x^{\beta}\chi^{\dagger}(v\cdot\Gamma)^{\dagger}-i\tilde{C}_{\mu}B_{\alpha}x^{\mu}\chi^{\dagger})\}\gamma_{0}\partial^{\alpha}\chi
\end{eqnarray*}
} where $\Delta$ encompasses the sign regarding the hermitian conjugation.
The remaining $\Omega$ functions appearing in\,\eqref{eq:identifications}
are obtained from the previous equations after the identifications
indicated in\,\eqref{eq:identifications}.

\end{document}